\documentclass[12pt,a4paper]{article}

\usepackage[a4paper,margin=1.5cm,footskip=1.3 cm]{geometry}
\usepackage{epsfig,latexsym}

\usepackage{graphicx}
\usepackage{amsmath}
\usepackage{amssymb}
\newcommand{\be}{\begin{eqnarray}}
\newcommand{\ee}{\end{eqnarray}}

\begin{document}
\begin{flushright}
CERN-PH-TH-2015-147\\
\end{flushright}
\vskip 24pt
\begin{centering}

{\bf \Large Observational Hints of a Pre--Inflationary Scale? } %\footnotemark[1]

\vskip 1 truecm

{\sc A.~Gruppuso${}^{\; a,b}$ \ and  \ A.~Sagnotti${}^{\; c,d}$} \\[20pt]

{${}^a$\sl\small INAF-IASF Bologna, \\
Istituto di Astrofisica Spaziale e Fisica Cosmica di Bologna, \\
Istituto Nazionale di Astrofisica, \\
via Gobetti 101, I-40129 Bologna, Italy} \\
%and \\ INFN, Sezione di Bologna, \\
%Via Irnerio 46, I-40126 Bologna, Italy \\ }
%e-mail: {\small \it gruppuso@iasbo.inaf.it}
\vspace{8pt}

{${}^b$\sl\small INFN, Sezione di Bologna, \\
Via Irnerio 46, I-40126 Bologna, Italy \\ }
e-mail: {\small \it gruppuso@iasbo.inaf.it}
\vspace{8pt}

{${}^c$\sl\small Department of Physics, CERN Theory Division\\
CH - 1211 Geneva 23, Switzerland \\ }
\vspace{8pt}

{${}^d$\sl\small
Scuola Normale Superiore and INFN\\
Piazza dei Cavalieri \ 7\\I-56126 Pisa \ Italy \\}
e-mail: {\small \it sagnotti@sns.it}

\vskip 2.5 truecm

%{\small (Dated: March 29, 2015)}

\end{centering}

%\maketitle

%\tableofcontents
%
%
%\newpage

\abstract{
\noindent
We argue that the lack of power exhibited by cosmic microwave background (CMB) anisotropies at large angular scales might be linked to the onset of inflation.
We highlight observational features and theoretical hints that support this view, and present a preliminary estimate of the physical scale that would underlie the phenomenon.}

%\date{\today}

\vskip 2 truecm

\begin{center}
{\sl Essay Written for the 2015 Gravity Research Foundation Awards for Essays on Gravitation.
Selected for Honorable Mention.}
\end{center}

\thispagestyle{empty}

\newpage

\noindent
{\bf Introduction.}
It is usually stated that the large scale Universe is fully characterized by the six parameters of the $\Lambda$CDM model \cite{Hinshaw:2012aka, Ade:2013zuv,Planck:2015xua}. Some features, however, are not well captured, and for instance anomalies occur at the largest CMB angular scales (see \emph{e.g} \cite{Copi:2010na}), although they are often regarded as mere curiosities. In this essay, we focus on one of these anomalies, the lack of correlation \cite{Hinshaw:1996ut,Spergel:2003cb,Bernui:2006ft,Copi:2006tu,Copi:2008hw,Efstathiou:2009di,Sarkar:2010yj,Gruppuso:2013dba,Copi:2013cya} and explain why, in our opinion, it deserves attention. The low variance anomaly \cite{Monteserin:2007fv,Cruz:2010ud,Gruppuso:2013xba,Ade:2013nlj} is a closely related observation \cite{Gruppuso:2013dba}, so that we shall use the terms ``lack of power'' and ``lack of correlation'' interchangeably.

%\section{Lack of power at large angular scales}

\vskip 0.5 truecm
\noindent
{\bf Lack of power at large angular scales.}
There is a lack of power, with respect to $\Lambda$CDM, in the two-point correlation function of CMB temperature anisotropies for angles larger than~$\sim~60^{\circ}$. This intriguing discrepancy was originally noted with COBE data \cite{Hinshaw:1996ut}, and was then confirmed by the WMAP team already in their first year release \cite{Spergel:2003cb}. In \cite{Bernui:2006ft} this feature was associated to missing power in the quadrupole. WMAP3 and WMAP5 data were then used to show \cite{Copi:2006tu, Copi:2008hw} that a lack of correlation occurs only in 0.03\% of the $\Lambda$CDM realizations. A subsequent analysis \cite{Efstathiou:2009di} confirmed the anomaly using WMAP5 data, and at the same time found, with a Bayesian approach, that the $\Lambda$CDM model cannot be excluded. WMAP7 data were taken into account in \cite{Sarkar:2010yj}, while
WMAP9 data were considered in \cite{Gruppuso:2013dba}, where the lack of correlation was studied against the Galactic masking.
{\sc Planck} 2013 and WMAP9 data were analyzed in \cite{Copi:2013cya}, which confirmed for this anomaly a significance at the level of $99.97\%$.

\begin{figure}[ht]
\centering
%\begin{figure}
\begin{tabular}{cc}
%\mbox{graphic} & \mbox{table} \\
\includegraphics[width=50mm]{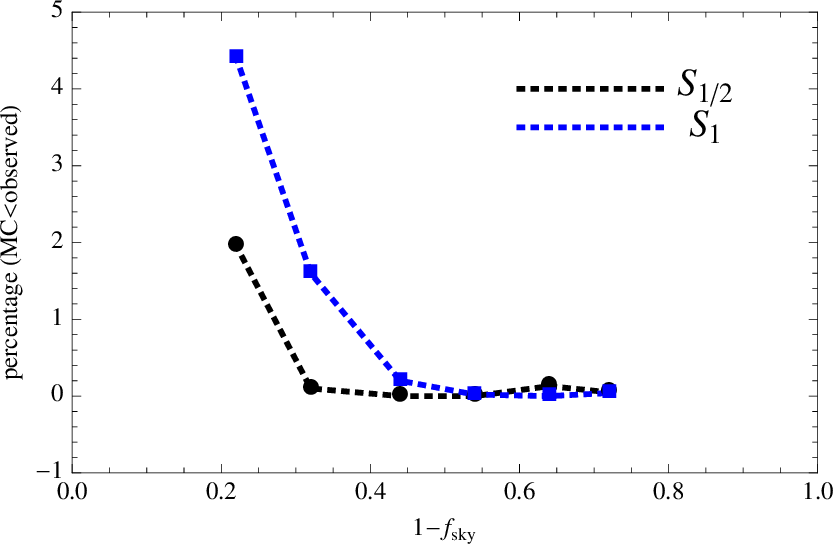} & \qquad\quad
\includegraphics[width=45mm]{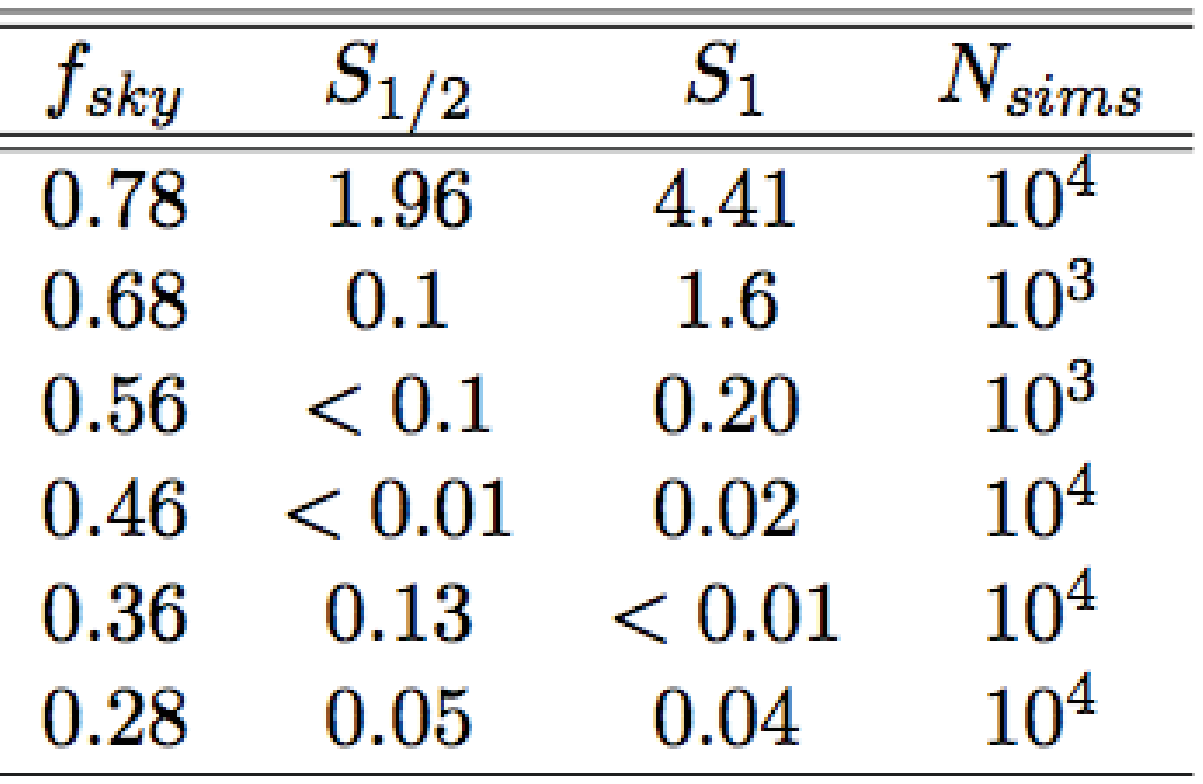} \\
%\mbox{\caption{\small Percentage anomaly (i.e. lower tail probability) of $S_{1/2}$ and $S_{1}$ versus the fraction of masked sky.}
%\label{figura1}} & \mbox{prova} \\
\end{tabular}
\caption{\small Percentage anomaly, or lower tail probability, of $S_{1/2}$ and $S_{1}$ versus the fraction of masked sky, for the standard mask and five extensions (left), and percentage probabilities to obtain smaller values than what observed by WMAP9 (right), from \cite{Gruppuso:2013dba}. $N_{sims}$ denotes the number of realizations.
Notice how the larger sampling variance accompanying lower values of $f_{sky}$ does not prevent this type of analysis.}
\label{figura1}
\end{figure}

%\subsection{Interesting features of the anomaly}
\vskip 0.5 truecm
\noindent
{\bf Interesting features of this anomaly.}
The two-point correlation function for CMB temperature anisotropies is defined as
\begin{equation}
C_{TT}(\theta) = \sum_{\ell \ge 2}^{\ell_{max}} \frac{(2 \ell +1)}{4 \pi} \  C_{\ell} \, P_{\ell}(\cos\theta) \, , \label{CTT}
\end{equation}
where the $P_{\ell}$ are Legendre polynomials and the $C_{\ell}$ are angular power spectrum (APS) coefficients.
In \cite{Gruppuso:2013dba}, $C_{TT}(\theta)$ was built via eq.~(\ref{CTT}) with a quadratic maximum likelihood estimator \cite{Tegmark:1996qt,Tegmark:2001zv,Efstathiou:2003tv,Gruppuso:2009ab,Molinari:2014wza}.
With it, one can compute the estimator~\cite{Spergel:2003cb}
\begin{equation}
S_{1/2} = \int_{\pi/3}^{\pi} d \theta \, \left( C_{TT} (\theta) \right)^2 \sin \theta \, ,
\label{s1su2estimator}
\end{equation}
together with its natural counterpart for the whole angular range~\cite{Gruppuso:2013dba},
\begin{equation}
S_{1} = \int_{0}^{\pi} d \theta \, \left( C_{TT} (\theta)  \right)^2 \sin \theta \, ,
\label{sfullestimator}
\end{equation}
which does not suffer from any a posteriori bias because it does not rest on an arbitrary choice of integration region. The estimators $S_{1/2}$ and $S_{1}$
were evaluated in \cite{Gruppuso:2013dba} for six different observed sky fractions $f_{sky}$, relying on realistic (signal plus noise) $\Lambda$CDM Monte Carlo (MC) simulations.
The Galactic masks were built extending the edges of the official kq85 WMAP mask
by 4, 8, 12, 16 and 20 degrees (see the table in fig.~\ref{figura1}).
These MCs were confronted with the values of $S_{1/2}$ and $S_{1}$ obtained from the ILC WMAP9 map, and
the WMAP9 observations displayed in general more compatibility with no correlation when the mask was enlarged.
For $S_{1/2}$, the probability to find a sky as the one observed by WMAP9 is $<0.01 \%$ for a Galactic mask preserving a sky fraction $f_{sky}=0.46$.
An anomalous probability below $0.01 \%$ for $f_{sky}=0.36$ is also obtained for $S_{1}$.
The percentages of compatibility are plotted in fig.~\ref{figura1}, which shows how the anomaly increases with the masked area.%\begin{figure*}

%\subsection{Why does the anomaly deserve attention?}
\vskip 0.5 truecm
\noindent
{\bf Why does this anomaly deserve attention?}
%We have convinced ourselves that
The observational facts reviewed in the previous section call for an explanation, since a statistical fluke remains possible only insofar as one accepts to be living in a very rare $\Lambda$CDM realization.
Why is this anomaly unlikely to originate from artifacts or spurious effects?
We can advocate at least three reasons for this.
To begin with, it is not natural to ascribe the lack of power to a foreground (\emph{e.g.} Galactic emission) that was not fully removed,
since Galactic and cosmological emissions are expected to be uncorrelated.
Therefore any residual foreground emission should increase rather than reduce the observed power \cite{Gruppuso:2007ya,Bunn:2008zd}.
Second, it is not natural to ascribe the effect to instrumental systematics, since two independent experiments, WMAP and {\sc Planck}, have observed the same features \cite{Copi:2013cya}.
Third, the anomaly is not only stable but its significance grows when the Galactic mask is increased, thus considering a smaller portion of the sky  \cite{Gruppuso:2013dba}.
This behavior, displayed in fig.~\ref{figura1}, is really remarkable, since the exclusion of regions close to the Galactic plane is in principle a very conservative choice.

\vskip 0.5 truecm
%\section{A new scale in Cosmology?}
\noindent
{\bf A new scale in Cosmology?}
Let us now elaborate on a possible fundamental origin for this effect.
At large angular scales the CMB anisotropy is governed by the Sachs-Wolfe (SW) effect, which is due to the gravitational potential at decoupling, \emph{i.e.} for $z \sim 1100$,
and by the Integrated Sachs-Wolfe (ISW) effect, which accounts for the more recent transition, at $z \sim 0.5$, between matter and dark energy domination.
The former effect is definitely more important, since the latter contributes less than $20 \, \%$ in terms of temperature anisotropies (see \emph{e.g.}~\cite{Ade:2013dsi,Ade:2015dva}).
Hence it appears reasonable, and even more conservative, to embed possible new mechanisms in the SW contribution.
%, which accounts for most of the power present at large angular scales.
%On the other hand, ISW modifications would seem more contrived, and moreover they ought to affect generically the pattern of high--multipole peaks, which is strongly constrained by observations.
On the other hand, ISW modifications would seem more contrived, since for one matter they should be fine-tuned in order not to affect the amount of dark energy, and thus the high--multipole peaks that are strongly constrained by observations.

The APS induced by the SW effect results from a Bessel--like transform \cite{cmbslow}, which projects on the sphere the primordial spectrum ${\cal P}(k)$ of scalar perturbations as
\begin{equation}
C_\ell \ = \ \frac{4\pi}{9} \ \int_0^\infty \frac{dk}{k} \ {\cal P} \big( k  \big) \, {j_\ell}^2 \big[ k \, (\eta_0 - \eta_{LS}) \big] \ ,
\end{equation}
where $\eta_0$ and $\eta_{LS}$ denote the conformal times at present and at last scattering.
In $\Lambda$CDM ${\cal P}(k)$ is parameterized as \cite{cm}
\begin{equation}
{\cal P}(k) \ = \ A \, k^{n_s -1}
\ ,
\label{primordial}
\end{equation}
in terms of an amplitude $A$ and a tilt $n_s-1$.

Lack of power at large angular scales is a typical manifestation of early departures from slow--roll, which follow naturally the emergence from an initial singularity. As explained in \cite{Dudas:2012vv,Kitazawa:2014dya}, when this occurs the power spectrum approaches in the infrared the limiting behavior
\begin{equation}
{\cal P}(k) = A \, \frac{k^{3}}{(k^2 + \Delta^2)^{2-n_s/2}} \ ,
\label{primordialmodified}
\end{equation}
which brings along a new physical scale $\Delta$. An infrared depression of the APS presents itself naturally in String Theory \cite{stringtheory}, in orientifold vacua \cite{orientifolds} with high--scale supersymmetry breaking \cite{bsb}. In these models a scalar field emerges at high speed from an initial singularity, to then bounce against a steep exponential potential before attaining an eventual slow--roll regime. The key ingredient is the steep exponential, whose logarithmic slope is predicted by String Theory \cite{dks,as13,fss}, and a number of exactly solvable systems provide explicit realizations of this peculiar dynamics \cite{fss}.
\begin{figure}[ht]
\centering
\includegraphics[width=80mm]{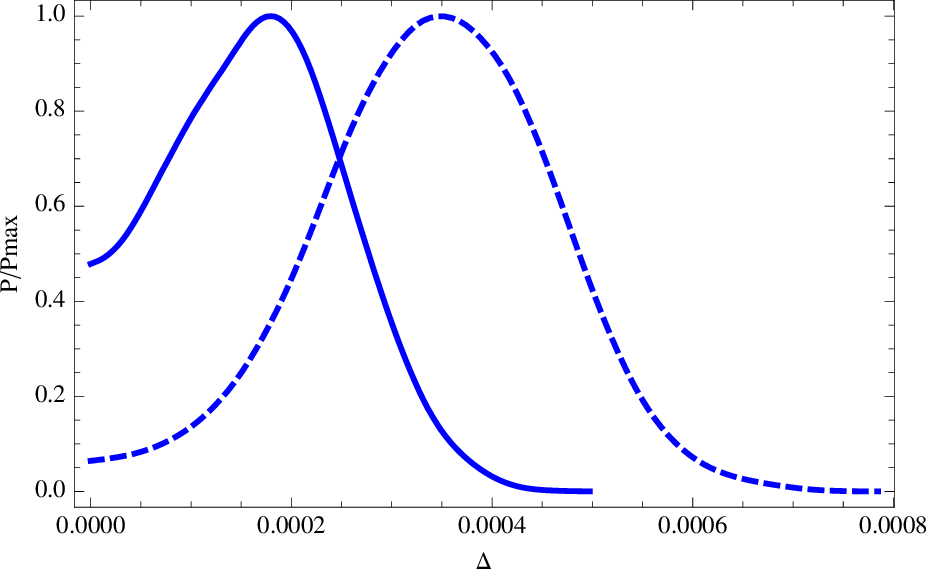}
\caption{\small Posterior probabilities of $\Delta$ (solid line for the standard mask with $f_{sky}~\simeq~90\%$, and dashed line for an extended mask with $f_{sky}~\simeq~40\%$).}
\label{figura2}
\end{figure}

Here we confine our attention to the simple deformation of eq.~\eqref{primordialmodified} of the almost scale invariant power spectrum \eqref{primordial}, leaving aside other secondary features introduced by a bounce \cite{Kitazawa:2014dya}. These include pre--inflationary peaks well separated from the limiting almost scale invariant profile \eqref{primordial}, to be contrasted with those accompanying ordinary transitions to slow roll, which occur next to it \cite{Destri:2008fj}.

The results of a preliminary Bayesian analysis extended to all standard cosmological parameters and based on {\sc Planck} data are shown in fig.~\ref{figura2},
where posteriors for $\Delta$ are given for two choices of the Galactic mask: the standard mask, with $f_{sky} \simeq 90 \%$ (solid line), and an extended mask with $f_{sky} \simeq 40 \%$.
The estimated value of the new scale $\Delta$ for the latter choice is
\be
\Delta = (0.340 \pm 0.115) \times 10^{-3} \, \mbox{Mpc}^{-1}  \, .
\label{Deltalargemask}
\ee
Interestingly, $\Delta$ in eq.~(\ref{Deltalargemask}) is found to differ from $0$ at $99\%$ C.L. and its magnitude appears reasonable, as we shall see shortly.
Moreover, in analogy with the lack of power anomaly, fig.~\ref{figura2} shows that the significance of this result increases sizably when a larger Galactic mask is used.

The string--inspired models of \cite{dks,Dudas:2012vv,Kitazawa:2014dya} rest on the energy scale of string excitations, which determines the scalar potential and lies typically a few orders of magnitude below the Planck scale. Inflation generally lasts ${\cal O}(100)$ times the corresponding time scale, and combining these results with
%well-established facts on the Universe from the radiation--dominated era, one can turn eq.~\eqref{Deltalargemask} into a primordial energy scale at the onset of inflation,
the standard evolution of the Universe, one can turn eq.~\eqref{Deltalargemask} into a primordial energy scale at the onset of inflation,
\be
\Delta^{Infl} \ = \ 4 \times 10^{15} \  e^{N-60} \times \sqrt{\frac{H_{Infl}}{100\, M_{Pl}}} \ \ {\rm GeV} \ .
\label{delta0}
\ee

This result depends on the number of $e$--folds, and demanding that $\Delta^{Infl}$ exceed slightly typical values of $H_{Infl}$ yields the inequality
\be
e^{N-60} \ \gtrsim \ \frac{M_{Pl} ({\rm GeV})}{4 \times 10^{15}} \ \sqrt{\frac{100 \ H_{Infl}}{M_{Pl}}}\ .
\ee
For an inflationary scale of about $10^{14}\ {\rm GeV}$ this implies the reasonable bound $N > 65$. Conversely, {\sc Planck} set the upper bound \cite{Ade:2015lrj}
\be
\frac{H_{Infl}}{M_{Pl}} \ < \ 3.6 \times 10^{-5} \ ,
\ee
and making use of this result in eq.~\eqref{delta0} yields
\be
\Delta^{Infl} \ \lesssim \ 2.4 \times 10^{12} \ e^{N-60} \ {\rm GeV} \ ,
\ee
which is again around $10^{14}\ {\rm GeV}$ for $N\simeq 65$.

In conclusion, the considerations in \cite{Dudas:2012vv,Kitazawa:2014dya}, inspired by String Theory, and in particular by the supersymmetry breaking mechanism of \cite{bsb} and the related cosmological dynamics of \cite{dks}, provided the original motivation for the present analysis. The resulting scenario would associate $\Delta$, and hence the corresponding primordial energy scale $\Delta^{Infl}$, to the onset of the inflationary phase, and as we have seen our estimates appear compatible with this interpretation. Collecting more information on the infrared tail of the APS might tell us something more definite about how an inflationary regime was originally attained.

%\newpage

%\section{Acknowledgments}

\vskip 1. truecm
%\section{A new scale in Cosmology?}
\noindent
{\bf Acknowledgments.}
%\vskip 0.5 truecm
\noindent
We are very grateful to N.~Kitazawa, N.~Mandolesi and P.~Natoli for several stimulating discussions and for an ongoing collaborations on these topics.
We are also grateful to M.~Lattanzi and L.~Pagano for discussions.
We acknowledge the use of computing facilities at NERSC (USA),
%of the HEALPix package \cite{gorski},
and of the {\sc Planck} Legacy Archive (PLA). We acknowledge also the use of the Legacy Archive for Microwave Background Data Analysis (LAMBDA), part of the High Energy Astrophysics Science Archive Center (HEASARC). HEASARC/LAMBDA is a service of the Astrophysics Science Division at the NASA Goddard Space Flight Center.
Work supported by ASI through ASI/INAF Agreement I/072/09/0 for the Planck LFI Activity of Phase E2, and by INFN (I.S. Stefi, FlaG).
%The work of NK was supported in part by the JSPS KAKENHI Grant Number 26400253.
AS in on sabbatical leave, supported in part by Scuola Normale Superiore.
The authors would like to thank the CERN Ph--Th Unit, Scuola Normale Superiore, the University of Ferrara and IASF--Bologna for the kind hospitality extended to them.

%We thank M. Lattanzi and L. Pagano for discussions.
%Some results presented in this paper are based on observations obtained with {\sc Planck} ({http://www.esa.int/Planck}), an ESA science mission with instruments and contributions directly funded by ESA Member States, NASA, and Canada. Moreover, we acknowledge the use of the Legacy Archive for Microwave Background Data Analysis (LAMBDA), part of the High Energy Astrophysics Science Archive Center (HEASARC). HEASARC/LAMBDA is a service of the Astrophysics Science Division at the NASA Goddard Space Flight Center. AG acknowledges support by ASI through ASI/INAF Agreement I/072/09/0 for the {\sc Planck} LFI Activity of Phase E2. AS in on sabbatical leave, supported in part by Scuola Normale Superiore and by INFN  (I.S. Stefi).
%The authors would like to thank the CERN Ph--Th Unit, Scuola Normale Superiore, the University of Ferrara and INAF--IASF Bologna for the kind hospitality extended to them while this work was in progress.

%

\end{document}